\documentclass[pdflatex,sn-mathphys-num]{sn-jnl}
\usepackage{graphicx}%
\usepackage{multirow}%
\usepackage{amsmath,amssymb,amsfonts}%
\usepackage{amsthm}%
\usepackage{mathrsfs}%
\usepackage[title]{appendix}%
\usepackage{xcolor}%
\usepackage{textcomp}%
\usepackage{manyfoot}%
\usepackage{booktabs}%
\usepackage{siunitx} %
\usepackage{algorithm}%
\usepackage{algorithmicx}%
\usepackage{algpseudocode}%
\usepackage{listings}%

\usepackage{xcolor}
\usepackage{subfiles}
\definecolor{codegreen}{rgb}{0,0.6,0}
\definecolor{codegray}{rgb}{0.5,0.5,0.5}
\definecolor{codepurple}{rgb}{0.58,0,0.82}
\definecolor{backcolour}{rgb}{0.95,0.95,0.92}
\theoremstyle{thmstyleone}

\theoremstyle{thmstyletwo}%
\theoremstyle{thmstylethree}%
\begin{document}
\title[Article Title]{An RRAM-based Hardware Implementation of a Radial Basis Function Neuron for Edge Classifiers}
\author*[1]{\fnm{Georgios} \sur{Papandroulidakis}}\email{gpapandr@ed.ac.uk}
\author[1]{\fnm{Shady} \sur{Agwa}}\email{shady.agwa@ed.ac.uk}
\author[1]{\fnm{Themis} \sur{Prodromakis}}\email{t.prodromakis@ed.ac.uk}
\affil[1]{\orgdiv{Centre for Electronics Frontiers, Institute of Micro and Nano Systems}, \orgname{School of Engineering, The University of Edinburgh}, \postcode{EH9 3BF}, \country{UK}}
\abstract{The deployment of modern machine learning (ML) solutions on resource-constrained edge devices highlights implementation challenges. This is especially true for extreme edge applications that include safety-critical components, such as autonomous navigation tasks. This paper demonstrates an artificial neural network (ANN) design leveraging Metal-Oxide Resistive RAM (RRAM) -based Analogue Content Addressable Memory (ACAM) as an efficient hardware substrate for performing metric-based classification and online adaptation on the edge. The proposed design is based on a custom Template piXeL (TXL) cell used for building the ACAM module, where each TXL cell acts as a configurable receptive field neuron. These cells employ a Radial Basis activation function to calculate the distance of an input from the programmed receptive field. The TXL can be organised into dense arrays for calculating the distance of a high-dimensional input against all stored prototypes, effectively performing fast and energy efficient similarity search. This hardware engine enables on-the-fly learning, where the receptive field parameters can be tuned to track domain shift. Through simulation of the proposed TXL-RBF classifier we can achieve 89.1\% accuracy on the MNIST dataset while consuming 185fJ per cell per operation when operating at 10MHz.}
\maketitle
In recent decades, researchers and engineers are actively rethinking computer architecture to achieve energy-efficient edge computing applications. This is particularly true for computationally-expensive algorithms such as modern Machine Learning (ML) and for applications that require strict specifications management due to being deployed in resource-constrained applications, such as edge computing cases \cite{Liu2021, prosensing_tinyml_9t4r}. The deployment of various Deep Neural Networks (DNNs) models on resource-constrained edge devices represents a significant frontier in modern computing \cite{Carmichael2018, Schlegel2015, Nandakumar2020}.  \par
Furthermore, the practical deployment of these models on the edge is impeded in many cases by their difficulties in efficiently handling outliers and out-of-distribution (OOD) inputs in real time \cite{fazaeli2025hard, gong2025augmented}. To mitigate these problems, robust on-device adaptation is required \cite{yang2025oodd, gong2025augmented, wang2025decoupled}. This is increasingly more important as the recent wide adoption and deployment of ML-driven autonomous edge systems leads to more frequent encounters with real-world data that differ from available training distributions. This phenomenon, known as domain shift, arises from variations in environmental conditions, sensor noise, or changes in data patterns over time, causing significant accuracy degradation in pre-trained models that have been deployed for inference in said applications. To ensure robustness and reliability, these edge ML models must adapt efficiently in-place to new data distributions without requiring additional resources or data annotation. Meta-learning is a ML framework aimed at introducing learning-to-learn (L2L) capabilities in Artificial Neural Networks (ANNs) \cite{son2024meta}. Meta-learning techniques provide solutions in the form of learning strategies, initialisation parameter methods, optimization procedures, or network architectures that facilitate rapid generalisation of existing ML capabilities to new target distributions \cite{javed2019meta}.  \par
In recent years, to alleviate the energy dissipation and latency due to von Neummann's memory wall bottleneck \cite{Papandroulidakis2014, Papandroulidakis2016}, considerable attention has been directed towards developing edge ML computing paradigms using emerging memory technologies and analogue in-memory computing (IMC) \cite{foster2025rram, yihan_associative_tcasi}. Novel multi-bit memory technologies, such as Resistive RAM (RRAM), can be effectively used to provide a flexible memory-centric computing substrate that has the potential to accelerate cornerstone ML operation, such as massively parallel dot-product operations \cite{Papandroulidakis2019, Xia2019}. Hybrid RRAM-CMOS circuits and systems are capable of performing analogue computing that encompasses high-level computational dynamics of biological neurons \cite{Serb2018}. Furthermore, systems implementing novel RRAM-based IMC classification engines have found applications in many different areas of novel ML hardware design, such as Content Addressable Memory (CAM) \cite{Chen2014, Zheng2017}. CAM and their analogue counterparts, Analogue CAM (ACAM), are specialized hardware memory architectures designed to perform parallel, in-memory similarity searches \cite{Ni2019, vardar202528nm, huang2025energy, Agwa2023_3T1R_ACAM, manea2025gain, wang2025decoupled, quan2024dual, pedretti2025solving}.  \par 
RRAM-based ACAMs store data as analogue conductance levels, by adjusting the RRAM device configurations, and processes inputs as analogue signals \cite{pedretti2025solving, Li2020_HP_ACAM, yihan_associative_tcasi, rbf_9t4r_iscas}. By exploiting the physics of the RRAM-CMOS ACAM cell, these systems can naturally calculate distance metrics \cite{Zhang2021, rbf_9t4r_iscas}. A recent example of RRAM-based ACAM can be found in the Template piXeL (TXL) technology, as described in \cite{Serb2018, foster2025rram, 9t4r_tcasi, rbf_9t4r_iscas}. TXL arrays have shown improved efficiency compared to alternative state-of-the-art ACAM cell technologies and competitive results in systems-level distance calculations \cite{9t4r_tcasi, rbf_9t4r_iscas}. \par 
In this work, we are proposing an edge in-silico classifier leveraging the RRAM-based Template piXeL (TXL) technology for metric-based prototypical memory classification and online adaptation at the edge. The TXL cell is employed as an analogue receptive field primitive used to build up associative memory arrays for low-latency and energy efficient similarity search operations. Each TXL matchline functions as a novel Radial Basis Function (RBF) -based artificial neuron with a programmable high-dimensional receptive field. This work presents the TXL array as a hardware-native prototype memory for edge classification. Each row stores a prototype, each column corresponds to an input feature, and the matchline response provides a similarity score. The same similarity-search operation supports classification, reliability estimation, and primitive prototype-level adaptation. The prototype adaptations used in TXL arrays represents an important step towards continual learning at the edge. Circuit-level simulations and energy estimation alongside a Python-based model and testing clearly illustrate the similarity search operation and the prototype adaptation operations. \par
\section*{Results}\label{sec:results}
\subsection*{RRAM-CMOS TXL Primitive} \label{results_subsec:txl__cell_primitive}
RRAM devices can be used to alter the effective voltage threshold of CMOS inverters when configuring artificial neurons \cite{Serb2018}. When we add $R_{M}$ in series with the pMOS source and $R_{B}$ in series with the nMOS source, the effective gate to source voltages ($V_{SGp}$ and $V_{SGn}$, respectively) are reduced due to the voltage drop across the source degeneration RRAM devices. Assuming a switching current $I_s$ flows through the stack, for nMOS devices we have: $V_{GSn} = V_{in} - I_s R_{B}$, while for pMOS devices we have: $V_{SGp} = (V_{DD} - V_{in}) - I_s R_{M}$. To find the new switching threshold $V_{TH}$ of the source degenerated CMOS gate, we set the square roots of the current equations (proportional to the over-drive voltages) equal to each other, adjusted by the MOSFET-based strength ratio $k_r$, as shown in Eq. \ref{vth}.  \par
\begin{equation}
    V_{TH} = \underbrace{\frac{V_{tn} + k_r\times(V_{DD} - |V_{tp}|)}{1 + k_r}}_{V_{TH}^{(0)}} + \frac{I_s\times(R_{B} - k_r R_{M})}{1 + k_r}
    \label{vth}
\end{equation}
$V_{TH}^{(0)}$ is the nominal voltage threshold value determined by the CMOS technology and configuration. Increasing $R_{B}$ (the nMOS source resistor) pushes the threshold higher. This is because the nMOS needs a higher input voltage to overcome the degeneration and become conductive enough to match the pMOS current. Increasing $R_{M}$ (the pMOS source resistor) pulls the threshold down. The equation in Eq. \ref{vth} demonstrates that the trigger voltage is a direct function of the resistance ratio $\Gamma_{rr} = R^{PU}_{M}/R^{PD}_{B}$. By programming the RRAM devices to specific ratios, the hybrid inverter can be tuned to any arbitrary analogue trigger point within the effective match range. The effective match range is defined by both the CMOS and RRAM technologies employed.  \par
\subsection*{RRAM-CMOS Radial Basis Function Neuron}\label{results_subsec:txl_cell_analysis}
\begin{figure*}[t!]
	\centering
	\includegraphics[width=\linewidth]{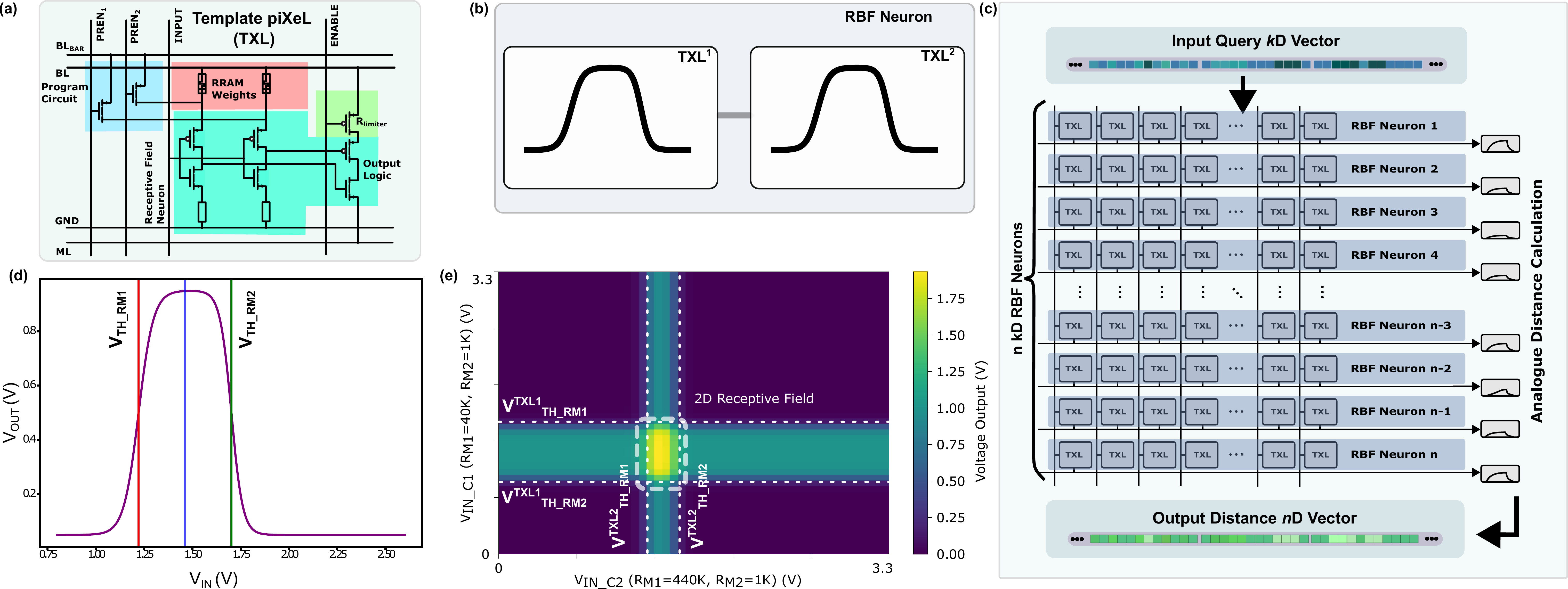}
	\caption{\textbf{A RRAM-based Scalable Reconfigurable silicon Radial Basis Function Neuron}: \textbf{(a)} A TXL cell implements effectively receptive field emulation neuron that can detect the input and fire only when the input falls within a specific matching window. The matching window is defined by the two RRAM devices $R_{M1}$ and $R_{M2}$ \cite{9t4r_tcasi}. \textbf{(b)} Interconnecting two such TXL-RBF circuits can be used to emulate higher dimensional receptive field, such as a 2D receptive field by interconnecting 2 TXL cells through their outputs. \textbf{(c)} In a similar manner, the interconnection of multiple TXL circuits can be used to emulate a neuron capable of capturing higher dimensional inputs, enabling an efficient similarity search of a high-dimensional query vector through a row of TXL cells. If we organise the TXL circuits in a $k^{columns}\times n^{rows}$ array we have $n$ $k-$dimensional neurons. Effectively, each row with interconnected outputs implements a $k-$dimensional neuron. The TXL matchline response can be interpreted as a hardware-friendly approximation to a Radial Basis Function -based similarity function. Each matchline stores through its RRAM-based receptive field weights a different class prototype. By using the massively parallel similarity search of the array, a set of similarity scores for every stored prototype given a single query vector can be calculated. Due to the RRAM-CMOS design of the TXL array, the module is compatible in a fully analogue dataflow with both input and output being represented by a set of analogue voltages. Each matchline responds with an analogue value that is linearly proportional with the level of match between the input and its stored internal memory. Thus, efficiently encoding the similarity score for this class prototype. \textbf{(d)} An example cell-level circuit response of the proposed TXL where a RBF is defined based on the two matching window thresholds. Each threshold is controlled by a resistive ratio. \textbf{(e)} An example matchline-level response when two TXL cells are interconnected. The cumulative effect is akin to a 2D receptive field.}
	\label{fig:figure_1}
\end{figure*}

The proposed TXL circuit, shown in Fig. \ref{fig:figure_1}(a), is based on two source degeneration -based RRAM-CMOS inverters. The two inverters are programmed to specific threshold values ($V_{TH}^{lower}$ and $V_{TH}^{upper}$) that together define a matching window. Each threshold value is implemented by configuring appropriately the conductance values of the programmable RRAM devices in the pull-up ($R^{PU} = R_{M}$) and pull-down ($R^{PD} = R_{B}$) networks. The effective behaviour of the TXL cell emulates a form of analog receptive field artificial neuron implementing a metric-based Radial Basis function (RBF). TXL can be organised into dense arrays to implement an effective high-dimensional receptive field neural network capable of performing similarity search operations. In Fig. \ref{fig:figure_1}(a) a 1D receptive field, Fig. \ref{fig:figure_1}(b) a 2D receptive field, and Fig. \ref{fig:figure_1}(c) a kD receptive field is shown. These networks are effectively implementing a RBF-based network capable of performing classification via similarity search operations. Example receptive fields based on TXL cell response are shown in Fig. \ref{fig:figure_1}(d) for a single cell output and Fig. \ref{fig:figure_1}(e) for the combined response of two cells connected to the same matchline. \par
In the programming scheme followed by our approach, $R_{M}$ represents the configurable RRAM source degeneration while the $R_{B}$ represents a fixed biasing resistive element that is used to implement the resistive ratio \cite{9t4r_tcasi}. This enables a more efficient programming scheme that requires minimal additional MOSFET circuitry per cell. The cell can be configured between two states, the 7T4R configuration during inference and two 1T1R configuration during training, where $R_{M1}$ and $R_{M2}$ are separately accessed and programmed to the appropriate conductance states. The TXL cell's matching window is characterized by a centre voltage $V_C$, shown in Eq. \ref{vc}, and a width $\Delta V$, shown in Eq. \ref{deltav} which are functions of the ratio pair $(\Gamma_{rr1}, \Gamma_{rr2})$. The voltage thresholds that define the matching window per cell are defined as shown in Eq. \ref{vth_final}. \par
\begin{equation}
    V_C = V_{TH}^{(0)} + \frac{I_s R_B}{1 + k_r} \left(1- k_r \times\frac{\Gamma_{rr, high} + \Gamma_{rr, low}}{2} \right)
    \label{vc}
\end{equation}
\begin{equation} 
    \Delta V = \frac{I_s R_B k_{r}}{1 + k_r} \left( \Gamma_{rr, high} - \Gamma_{rr, low} \right)
    \label{deltav}
\end{equation}
\begin{equation}
    V_{TH}^{lower} = V_C - \frac{\Delta V}{2},
     \qquad V_{TH}^{upper} = V_C + \frac{\Delta V}{2} 
     \label{vth_final}
\end{equation}
\subsection*{TXL-based Memory Retrieval and Adaptation} \label{results_subsec:txl_array_retrieval_adaptation}
TXL cells are organized into a dense array where each column (or bitline (BL)) of the array corresponds to a specific dimension of a $D$-dimensional feature vector $\mathbf{x} = (x_1, \dots, x_D)$, and each row (or matchline (ML)) of the array represents a unique $D$-dimensional prototype $m$. Each TXL cell is represented the statistical attributes $(\boldsymbol{\mu}_{m,D}, \sigma_{m,D})$, that are encoded to the RRAM parameters as $V_{C} = \boldsymbol{\mu}_{m,D}, \Delta V =  2 \times \sigma_{m,D})$, through Eq. \ref{vc}, \ref{deltav}, \ref{vth_final}. The TXL array enables per-feature similarity calculation and then accumulates the total similarity for each prototype through in-memory analogue signal processing. More specifically, the output is connected to a single node per prototype due to the connectivity of the TXL array, enabling a natural charge accumulation of each cell to the matchline output via Kirchhoff's law of currents. The charge accumulation naturally encodes the similarity score for each matchline, with the similarity score being the inverse distance of the query against the prototype. \par
\begin{figure*}[t!]
	\centering
	\includegraphics[width=\linewidth]{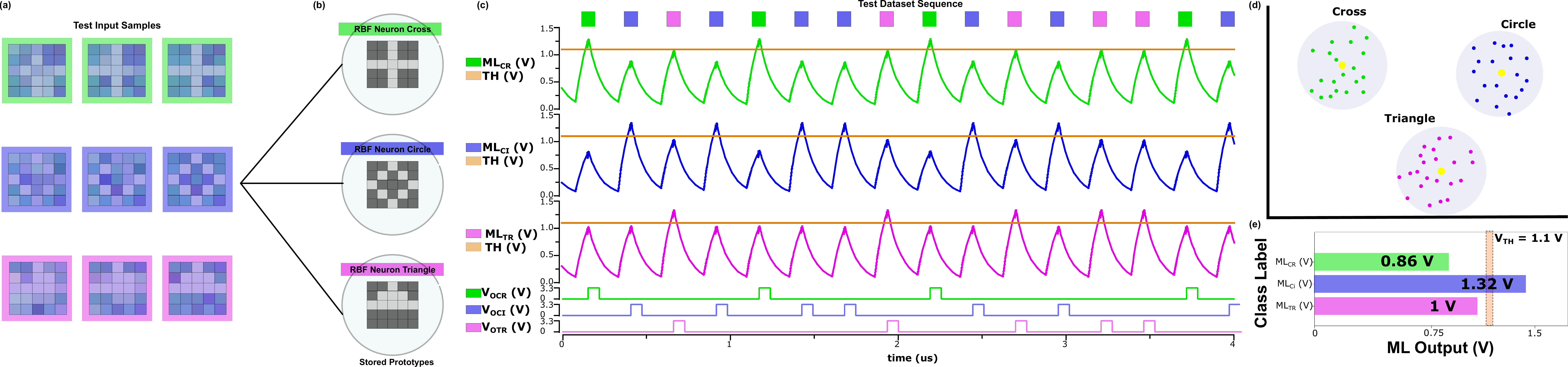}
	\caption{\textbf{Demonstration of RRAM-CMOS silicon TXL RBF neuron}: \textbf{(a)} Examples from the different symbols that are used for training. The dataset employed for these tests is an artificial symbol dataset comprised of three geometric shapes. \textbf{(b)} Each class generates a representative prototype that is stored in the TXL array as a prototype. \textbf{(c)} Spectre-SPICE simulation results for an example implementation of the proposed TXL neuron as a small network capable of storing the serialised prototypes and perform in-situ similarity search. The top three traces showcase the analogue output of the array when a series of binarised geometric symbols are shown as queries to the test network. \textbf{(d)} The proposed network effectively performs a metric-based clustering operation. The clusters are defined based on the formed representative prototypes stored in the TXL-based memory and the in-distribution threshold hyperparameters of the model that defines the maximum association of an input with a cluster. The representative prototypes are formed by taking the average of all training sample embeddings. \textbf{(e)} For each decision in the showcase example simulation, the full similarity search and classification operation is performed within the TXL-based array.}
	\label{fig:figure_2}
\end{figure*}

As shown in Fig. \ref{fig:figure_2}, TXL arrays can be used to perform parallel similarity search in-place, providing a template-matching calculation with $O_{(1)}$ complexity \cite{9t4r_tcasi}. For demonstration purposes, simple geometric shapes of $5\times5$ size are generated to form an artificial symbols dataset. For each class an average representative prototype is generated through averaging and binarisation of multiple versions of these symbols. Different versions of each symbol are generated by introducing a specific level of noise per input sample. As part of the artificial dataset pre-processing, the symbols are spatially converted to 1D vectors. The binarised prototypes are stored as parameters to a prototype $N_{columns}^{features}\times M_{rows}^{prototypes} = 32\times48$ TXL array \cite{9t4r_tcasi}. The array employs two different TXL cell configurations to map the binarised prototypes by using a $R_{M1}=10K\Omega, R_{M2}=10K\Omega$ and a $R_{M1}=100K\Omega, R_{M2}=300K\Omega$ with matching window centres at $1.9V$ and $1.6V$, respectively, and showcasing no overlapping matching window activity. The resistance of the pull-down networks is fixed for all configurations to $R_B = 200k\Omega$. The analogue response of each matchline is being subjected to threshold matching through integrated sense amplifiers with $V_{REF}^{SA} = 1.1V$, resulting in a one-hot encoding for the final classification output \cite{rbf_9t4r_iscas}. \par
In Fig. \ref{fig:figure_2} an example inference operation through the prototype TXL array is shown. Example samples from the 3 shapes available in the artificial dataset are shown in Fig. \ref{fig:figure_2}(a). In the example symbol dataset multiple samples of cross, circle and triangle symbols are generated. A randomised noise mask is added per sample to implement the training and testing datasets. The samples per class are aggregated, averaged and binarised to create representative templates that can be used to map the main form of each symbol. The templates comprise effectively statistically driven prototypes of each class and are stored in separate TXL rows. The 2D prototypes are shown in Fig. \ref{fig:figure_2}(b). The TXL array requires the transformation of the 2D prototypes into 1D vectors to store in each TXL matchline. An example SPICE-Spectre circuit-level simulation is shown in Fig. \ref{fig:figure_2}(c). The three upper traces showcase the analog matchline voltage charging and discharging during the reset-evaluation cycle of repeated inference operations. Each inference operation performs a similarity search of a given input query vector with all stored prototypes simultaneously. The sequence of this test inference is shown as coloured rectangles above the traces, where green, blue and pink encode cross, circle and triangle test samples, respectively. The circuit is calibrated by adjusting the voltage reference of the output sense amplifiers to provide a unique classification output for each test sample. This becomes apparent when observing the charging characteristics of all matchlines per test sample. For a given sample only one out of three matchline charges beyond the voltage reference level signifying a match event. The charging behaviour is based on the accumulation of multiple TXL cells firing due to attributes of the test sample falling within their respective matching window. In this work, the matchline output is interpreted as a prototype similarity score. A higher score indicates that the input feature vector falls strongly within the programmed feature-wise acceptance windows of a stored prototype. This is mapped in hardware through the physical quantity of charge accumulation per matchline. The lower three traces showcase the digitised TXL array response, as part of the sense amplifier output integrated within the array. An approximate clustering plot is shown in Fig. \ref{fig:figure_2}(d), with each training sample being used to define a cluster region per class. An example similarity search operation is shown in Fig. \ref{fig:figure_2}(e). The analog output per matchline is evaluated as the similarity level of the input query of the inference operation with the specific matchline. By surpassing an arbitrary defined threshold, a specific matchline indicates a match event that leads to the classification of the specific input query. \par
The configuration of TXL arrays as a configurable form of RRAM-based prototypical memory module, enables the fast and computationally inexpensive adaptation of the store prototypes. The adaptation refers to the reprogramming of the appropriate RRAM devices to adjust existing prototypes or introduce new prototypes alongside the existing ones to mitigate domain drift issues encountered during deployment. Before any adaptation can take place, the array categorizes the input according to its similarity with the available prototypes. More specifically, it checks whether the inverse similarity, thus distance $d$, for the best matching prototype is below two statistically-defined thresholds, $\tau_{IDO}$ and $\tau_{OOD}$. The two thresholds implement a form of gating mechanism that provides a transparent method to assess he trustworthiness of the classification output of the TXL array. The thresholds are calculated based on the global training distribution statistics that define the per-prototype matching similarity score distributions. These parameters can be efficiently mapped as reference voltages of the implemented sense amplifiers of the array. The logic that defines the reliability of each similarity search operation is shown in Table \ref{tab:adaptation_policy}. \par

\begin{table}[ht]
    \centering
    \caption{Adaptation Policy}
    \label{tab:adaptation_policy}
    \begin{tabular}{ll} 
        \toprule
        \textbf{Best Match Conditions} & \textbf{Interpretation} \\ 
        \midrule
        $d \leq \tau_{IDO} $        & Reliable in-distribution input, no adaptation \\ 
        $\tau_{IDO} < d \leq \tau_{OOD} $        & In-distribution outlier, known prototype adaptation \\ 
        $d > \tau_{OOD} $         & OOD candidate, allocate new prototype \\ 
        \bottomrule
    \end{tabular}
\end{table}

The input is considered reliable (fully in-distribution), when it is close to a known existing prototype centre and $d$ is below both thresholds $\tau_{IDO}$ and $\tau_{OOD}$. For reliable inputs no adaptation is performed to any stored prototype. If the input has $d$ higher than $\tau_{IDO}$ but below $\tau_{OOD}$ then it is categorised as unreliable (an in-distribution outlier). This could be the case for an input that is recognisable as belonging to a known class but lies toward the periphery of its assigned cluster, suggesting gradual drift. In this case, the specified closest prototype is adapted towards including the input and categorising it as reliable. Finally, if the input has $d$ higher than $\tau_{IDO}$ and $\tau_{OOD}$ then it is categorised as out-of-distribution (OOD). In this final case, the input effectively cannot be matched with any stored existing prototype. Instead, it is considered an input from a novel previously unseen class. To enable future classification of this input, a new representative prototype needs to be generated and added alongside existing prototypes. For the proof-of-concept test shown here, supervised adaptation is used, thus the label is available to the classifier. \par
\begin{figure*}[t!]
	\centering
	\includegraphics[width=\linewidth]{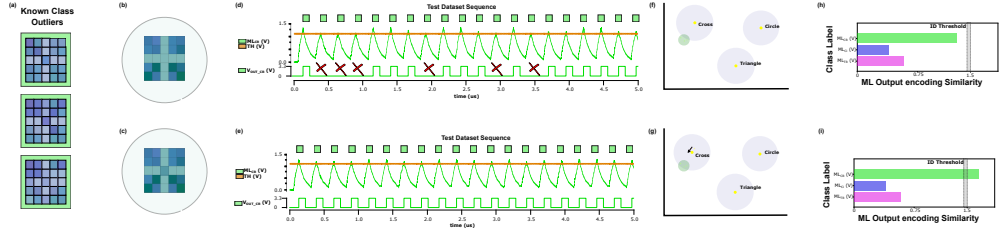}
	\caption{\textbf{In-distribution adaptation of known class prototypes}: \textbf{(a)} When a few instances of a known class are presented to the TXL classifier with the best matching distance $d$ is above $\tau_{IDO}$ but below $\tau_{OOD}$, then a prototype adaptation mechanism is triggered. \textbf{(b)} The weighted average outlier template is applied to the existing prototype. \textbf{(c)} A new prototype, adapted to the outliers weighted features, is stored in TXL array. \textbf{(d)} SPICE simulations of the initial cross template against outlier cross samples \textbf{(e)} SPICE simulations of the adapted cross template with the same series of outlier samples as before. The system performed a test-time adaptation of the closest class prototype. This enabled the reshaping of the prototype to more accurately capture the previously known samples of the class as well as the new outlier samples. \textbf{(f)} The formed clusters before adaptation. \textbf{(g)} The formed clusters after adaptation. \textbf{(h)} The distance calculation before adaptation. \textbf{(i)} The distance calculation after adaptation. }
	\label{fig:figure_3}
\end{figure*}

An example adaptation operation for adjusting an existing prototype in the TXL array is shown in Fig. \ref{fig:figure_3}. In Fig. \ref{fig:figure_3}(a), a few-shot learning operation is initiated by the stimulation of the TXL array through specific noisy cross inputs. These specific samples are outliers and are close to the edge of the existing class cluster (as shown in the approximate clustering plot in Fig. \ref{fig:figure_3}(f)). By querying these samples the TXL outputs multiple misses (shown as red cross symbols on Fig. \ref{fig:figure_3}(d)). This shows that the TXL array cannot reliably classify the specific subset of test samples when compared with the pre-adaptation prototype (shown in Fig. \ref{fig:figure_3}(b)). The SPICE-Spectre circuit-level simulations of the pre-adaptation matchline response are shown in Fig. \ref{fig:figure_3}(d). A few-shot learning scheme is used to adjust the existing prototype into a new adapted prototype (shown in Fig. \ref{fig:figure_3}(c)). The few-shot learning follows a supervised clustering method. By testing the same outlier sequence, the TXL array can now correctly identify the outlier test samples. The adapted circuit-level simulation results are shown in Fig. \ref{fig:figure_3}(e). Example clustering plot and distance calculations for pre-adaptation and post-adaptation cross-prototype are shown in Fig. \ref{fig:figure_3}(f,h) and Fig. \ref{fig:figure_3}(g,i), respectively.  \par
The TXL array supports a primitive hardware-friendly method to continual learning through prototype update or row allocation. It enables local modification of the associative memory without full model retraining, but semantic labelling of new classes requires external supervision or confirmation. While in-distribution outlier adaptation allows existing prototypes to adapt to domain drifts, the out-of-distribution (OOD) adaptation policy handles the encounter of novel classes for a given classification task. TXL can adapt its encoded knowledge by introducing new memory entries that represent new prototypes, ensuring the system can expand its classification capabilities without removing or altering existing prototypes and thus, potentially avoiding catastrophic forgetting of existing knowledge. \par
\begin{figure*}[t!]
	\centering
	\includegraphics[width=\linewidth]{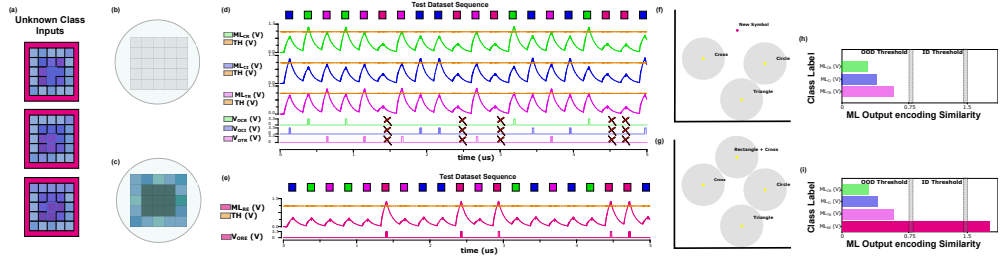}
	\caption{\textbf{Out-of-distribution adaptation of prototypical memory for unknown classes}: \textbf{(a)} When an input of a previously unseen class is received by the TXL classifier, the system can compare it to exiting stored prototypes and trigger an out-of-distribution (OOD) adaptation procedure. The OOD adaptation is triggered for the case when all computed distances are higher than the defined OOD threshold $\tau_{OOD}$). \textbf{(b)} An available knowledge prototype entry is selected to store the novel class detected based on on the OOD adaptation mechanism. \textbf{(c)} The new prototype is stored in TXL prototypical memory and can be recalled in future classification operations. \textbf{(d)} SPICE simulation showing the TXL response before OOD adaptation for the new class is applied. The red input samples represent samples of the rectangle symbols which was not part of the initial offline training. \textbf{(e)} SPICE simulation showing the new TXL neuron response after that compares the same input sequence adapted prototype. This new prototype has been added alongside the existing prototypes. \textbf{(f)} In the clustering operation the new input sample is being placed further away that the minimum distance required to be considered a in-distribution sample. \textbf{(g)} The new input is used to create a new prototype representation in the clustering space. \textbf{(h)} Distances below the OOD threshold before the new prototype integration. \textbf{(i)} Distance calculation with the newly added prototype. }
	\label{fig:figure_4}
\end{figure*}

An example operation for adapting the TXL array by introducing a new representative prototype is shown in Fig. \ref{fig:figure_4}. The adaptation is triggered by stimulating the TXL array with test samples from a class that was not part of the initial 3 main symbols dataset. More specifically, test samples from a fourth class, noisy inputs representing rectangle (with examples shown in Fig. \ref{fig:figure_4}(a)), are used to query the TXL array. A pre-adaptation circuit-level simulation showcase misses when queried with samples of this unknown class, as shown in Fig. \ref{fig:figure_4}(d). For the red test samples, as indicated by the test sequence at the top of the simulation traces, no existing prototype provides a reliable level of match for these out-of-distribution test samples. To adapt the TXL array with a new prototype, a supervised clustering method to quickly adapt TXL array with an additional prototype to represent the new symbol. An empty TXL row (shown in Fig. \ref{fig:figure_4}(b)) is populated with a representative prototype based on the average feature map generated by the new class input samples (shown in Fig. \ref{fig:figure_4}(c)). The adapted SPICE-Spectre simulation results are shown in Fig. \ref{fig:figure_4}(e), where for the same sequence of test inputs the newly added prototype fires when a red (rectangle) test sample is queried. The new prototype exists alongside the initial 3 prototypes and no changes to these initial prototype configuration have been applied. Example clustering plot and distance calculations for pre-adaptation and post-adaptation cross prototype is shown in Fig. \ref{fig:figure_4}(f,h) and Fig. \ref{fig:figure_4}(g,i), respectively. The new samples are used to create a new cluster and when queried the new matchline provides a good level of match compared to the pre-existing prototypes. \par
For calculating the approximate energy dissipation of our proposed classifier, we used Cadence Virtuoso post-layout simulations to access the behavioural response of the circuits and systems under test. For the back-end classifier, each template matching operation consumes $\SI{185}{\femto\joule}$ per cell. For a fixed-size TXL array, all prototype rows are searched in parallel, giving approximately constant search latency with respect to the number of stored prototypes within that array. Area and energy scale with the number of active rows and columns. The main characteristics of the prototype TXL IC, TXL9v1, are shown in Table \ref{tab:hardware_characteristics}. Assuming the use of all 32 feature lines (columns) and all 48 prototypes (rows), thus using the full $32\times48$ TXL prototype IC, then the approximate energy for the core TXL array is:
\begin{equation}
\begin{split}
    E_{\text{TXL}} = N_{\text{features}} \times N_{\text{prototypes}} \times E_{\text{cell}} = 32 \times 48 \times 185fJ = 0.284nJ
\end{split}
\end{equation}
\begin{table}[ht]
    \centering
    \caption{Hardware Characteristics of the RRAM-CMOS TXL ACAM}
    \label{tab:hardware_characteristics}
    \begin{tabular}{ll} 
        \toprule
        \textbf{Characteristics} & \textbf{TXL IC \cite{9t4r_tcasi}} \\ 
        \midrule
        CMOS Technology          & \SI{180}{\nano\meter} MOSFET (\SI{5}{\volt}) \\ 
        RRAM Technology          & $\text{TiN/HfON/TiN}$ \\ 
        Voltage Supply           & \SI{3.3}{\volt} \\ 
        Energy (per cell per op) & \SI{185}{\femto\joule} \\ 
        Latency (per search)     & \SI{100}{\nano\second} \\ 
        Area (of TXL array)      & \SI{1.13}{m\meter\squared} \\ 
        Capacity                 & $32^{columns} \times 48^{rows}$ array \\ 
        Cell Count               & $1536$ cells \\
        \bottomrule
    \end{tabular}
\end{table}
To test the performance of the proposed technology we have used the artificial symbol dataset as well as the MNIST dataset to test the training, inference and adaptation of TXL arrays. The artificial symbol dataset is used to illustrate the TXL memory-retrieval and adaptation mechanism under controlled conditions. MNIST is used as a more complex benchmark for the prototype-classification concept as implemented via TXL. These additional tests were perform using an approximate Python model for describing the main operational principles of TXL. To test the behaviour of the TXL array, we initially performed an offline supervised training for the known classes and tested the inference on this initial set of classes. After an example in-distribution adaptation (adaptation of an existing class) and new prototype introduction (adaptation of existing knowledge by introducing a new prototype via few-shot learning scheme) we tested the inference of the TXL array again. For the case of the example artificial symbol dataset, TXL can achieve 100\% accuracy for the initial offline training. After the adaptation we are able to still achieve also approximately 100\% accuracy. This showcases that the proposed TXL array can be used effectively for simple geometric shape recognition. In the case of more complex templates, such as the MNIST dataset, the TXL can achieve accuracies of 89.1\% pre-adaptation and 85.5\% post-adaptation, respectively, showcasing competitive results. This slight drop in accuracy can be tracked to the use only of the TXL array as classifier without any feature extraction or any linear projection before of after the array. Additionally, the MNIST dataset is downsampled to $7 \times 7$ size before use. The downsampling does not include any learnable feature extraction. Finally, for the case of the MNIST dataset that requires an input vector larger than 32 features ($7\times7 = 49$), multiple TXL arrays are assumed to work in parallel towards storing the full prototypes. The outputs were combined to provide the final classification. The MNIST Python test was performed as a behavioural extension of the TXL classification model using the stated downsampled image dimensionality. \par
\section*{Discussion}\label{sec:discussion}
In this work, we propose a novel RRAM-based hardware ML computing paradigm for metric-based classification and online adaptation at the edge. More specifically, RRAM-based Template piXeL (TXL) arrays are leveraged to implement a fast, energy-efficient classifier capable of performing similarity search in the analog data domain. We define the TXL cell as a custom configurable Radial Basis Function (RBF) artificial neuron, an analog circuit emulation of a receptive field neuron. Through dense TXL arrays we can implement a hardware-friendly similarity search engine via an In-Memory Computing (IMC) scheme \cite{lei2025radial, rbf_9t4r_iscas}. The TXL’s activation function maps to a weighted distance calculation of an input with its stored receptive field, enabling both metric-based classification and computationally efficient adaptation of the TXL prototypical memory \cite{wu2026review, chen2025knowledge}. The behavioural response of the prototype TXL array is shown via an artificial dataset to explain how the knowledge is stored, how the similarity search works and how the knowledge is adapted to match target domain distribution. \par
Memory-centric in-silico classifiers, such as the proposed TXL array, can operate within a limited power budget \cite{prosensing_tinyml_9t4r}, making it viable for resource-constrained environments like remote monitoring systems, wearable sensors, autonomous navigation, and streaming processing edge applications that require real-time classification. The reliance on distance-based classification provides an inherently explainable approach compared to the usually opaque feed-forward dense layers, as the decision is explicitly traceable to the proximity of a learned representative prototype stored as integrated knowledge in the TXL array. Furthermore, the adaptation of the TXL prototypical memory can provide robustness when the system is exposed to real-world target domain information. This feature is particularly valuable for applications such as autonomous navigation of vehicles, industrial anomaly detection systems, and other safety-critical applications, which frequently encounter unseen target domain stimuli that deviate from the original training domain \cite{ Halawani2021, li2025cim, yang2025end, agwa2025oisma}.
As part of future work, the deployment of the proposed TXL technology for practical edge computing paradigms is envisioned. This could be explored through the implementation of Test-Time Adaptation (TTA) custom architectures with TXL being deployed as a configurable prototypical memory \cite{yang2025oodd, gong2025augmented, wang2025decoupled}. Practical deployment will require calibration for RRAM variability, programming error, retention, endurance, and temperature-dependent drift. These effects may shift the programmed receptive fields and therefore affect classification thresholds and adaptation reliability. \par
\section*{Methods}\label{sec:methods}
\subsection*{RRAM \& CMOS Technology Integration} \label{methods_subsec:rram_cmos_technology}
In this work, we are employing $TiN/HfON/TiN$-based metal-oxide Metal-Insulator-Metal (MIM) material stack with layer thickness of approximately $50/5/50 nm$ \cite{tsiamis2026rapid}. The device showcase bi-polar behaviour with $\SI{1}{\volt} - \SI{3}{\volt}$ requires for programming purposes and up to $\SI{5}{\volt}$ for the initial electroforming operation. The reading voltages employed should not exceed $\SI{0.5}{\volt}$ to avoid any conductance state drifting. The devices showcase good analogue memory behaviour when stimulated with controlled pulse trains \cite{Stathopoulos2017a, Stathopoulos2018, tsiamis2026rapid}. For the purposes of this design, we are employing a commercially available $\SI{180}{\nano\meter}$ CMOS technology. We are using both $\SI{1.8}{\volt}$ and $\SI{5}{\volt}$ components from this technology. All main components of the core TXL array circuits and systems, forming a RRAM-CMOS ACAM macro module, are implemented on $\SI{5}{\volt}$ MOSFET devices, since the RRAM-related circuitry requires higher voltages to handle the electroforming and programming processes \cite{tsiamis2026rapid}.  \par
RRAM devices are integrated through a custom in-house Back-End-Of-Line (BEOL) integration process to the CMOS substrate \cite{tsiamis2026rapid}. Through this design methodology the RRAM-CMOS implementation is completed with two steps, with the first being the tape-out of the CMOS computing substrate and the second being the in-house integration of RRAM devices. The design layouts assumed the above method of implementation according and followed the design rules for the CMOS and RRAM devices. More details on the BEOL integration method followed for the implemented TXL prototype IC can be found in \cite{tsiamis2026rapid}.  \par
\subsection*{Circuit-Level Simulations and Energy Calculations}\label{methods_subsec:simulation_and_energy}
Cadence Virtuoso and a fully analog design and simulation flow is used to perform any circuits and systems -level simulations in this work. The SPICE simulations are post-layout and showcase the behaviour of the proposed TXL cells and arrays with parasitic resistances and parasitic capacitances according to the $\SI{180}{\nano\meter}$ CMOS technology employed for the RRAM-CMOS design. An in-house data-driven RRAM model based on experimental measurements of real RRAM devices was used to simulate the behaviour of the RRAM devices \cite{Messaris2017, yihan_associative_tcasi}. \par
The TXL implementation showcase competitive energy efficiency compared to other state-of-art ACAM technologies of approximately $\SI{185}{\femto\joule}$ per similarity search operation per cell \cite{9t4r_tcasi}.  \par
\begin{figure*}[h!]
	\centering
	\includegraphics[width=\linewidth]{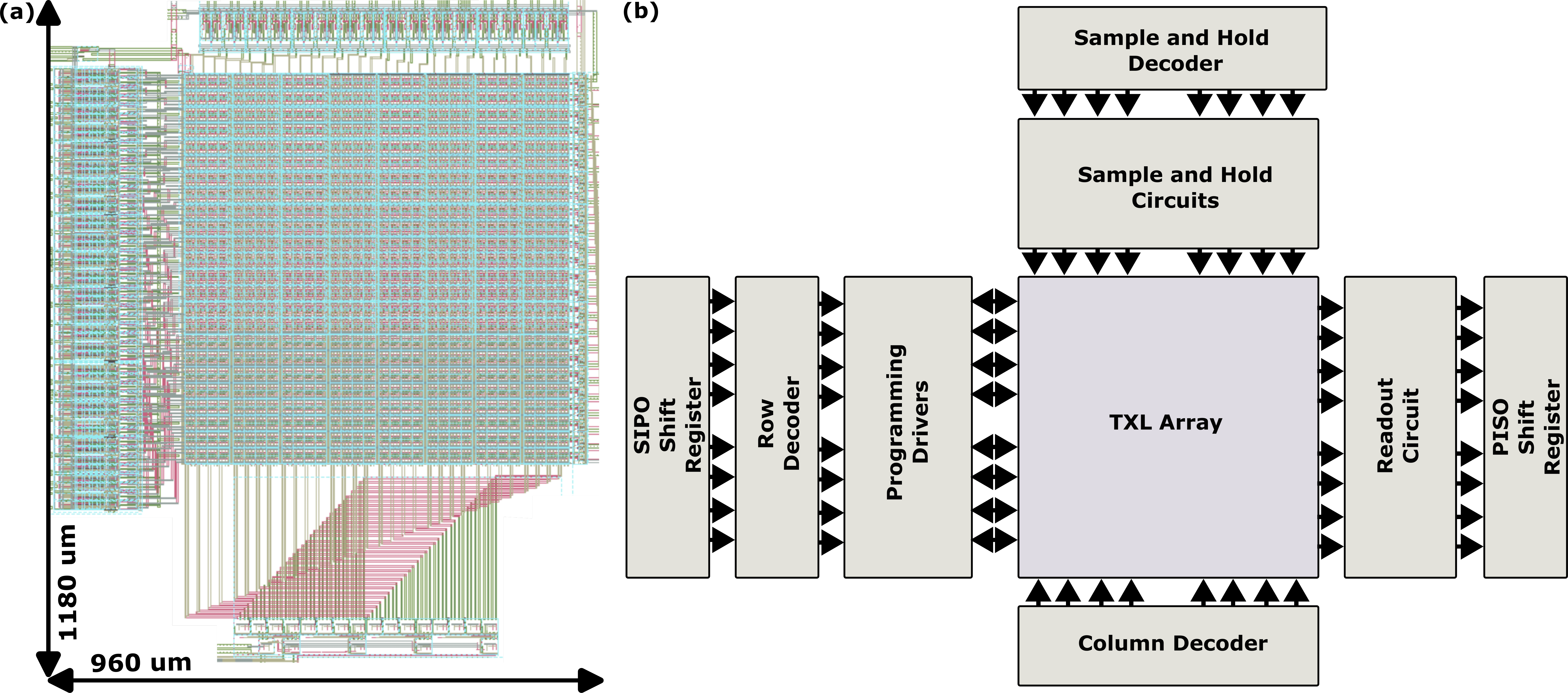}
	\caption{\textbf{TXL Array Implementation Prototype TXL IC \cite{9t4r_tcasi}}: \textbf{(a)} Layout of the TXL9v1 core array implemented in $\SI{180}{\nano\meter}$ $\SI{5}{\volt}$ technology. The main TXL core array is $\SI{1.13}{m\meter\squared}$ with the full IC having an area of $\SI{3.8}{m\meter\squared}$ \cite{9t4r_tcasi}. \textbf{(b)} Block diagram of the implemented prototype TXL IC showcasing the main components of the array. The TXL array and the peripherals for programming and reading the memory and performing the single step similarity search are implemented with $\SI{5}{\volt}$ MOSFET devices. Shift registers for custom serialisation and deserialisation control and output readout of the main core circuits is introduced to enable a smaller pin count for the full prototype TXL IC. }
	\label{fig:txl_prototype_ic}
\end{figure*}

\subsection*{Prototype Representation and the RRAM Conductance Encoding} \label{sec:rram_programming_method}
A linear mapping between the receptive field of each prototype statistical representation and the physical resistance state of the corresponding RRAM device pair per TXL cell is followed for the programming scheme for the SPICE-Spectre simulations. For each receptive field in a prototype, mean voltage $\mu_{m,i}$ and variance $\sigma_{m,i}$ is encoded per feature. The RRAM devices require to be configured as shown in Eq. \ref{rm1_encoding} and Eq. \ref{rm2_encoding} to encode the required mean and variance per feature. 
\begin{equation}
    R_{M1,m,i} = \frac{R_B}{k_r} - A\,(\mu_{m,i} - \sigma_{m,i} - V_{TH0})
    \label{rm1_encoding}
\end{equation}
\begin{equation}
    R_{M2,m,i} = \frac{R_B}{k_r} - A\,(\mu_{m,i} + \sigma_{m,i} - V_{TH0})
    \label{rm2_encoding}
\end{equation}
For Eq. \ref{rm1_encoding} and Eq. \ref{rm2_encoding}, the hardware-related parameters $A = (1+k_r)/(I_s \cdot k_r)$, $k_r = \sqrt{\beta_p/\beta_n}$, $I_s$, and nominal voltage threshold $V_{TH0} = (V_{tn} + k_r(V_{DD}-|V_{tp}|))/(1+k_r)$ are based on the underlying CMOS technology utilized for the prototype TXL IC. \par
\subsection*{Analog Distance Computation via RRAM-enabled Physical Computing} \label{sec:rram_physical_computing}
The TXL cell activation function approximates a normalised RBF receptive field:
\begin{equation}
    g_{m,i}(x_i) = \exp\!\left(-\frac{(x_i - \mu_{m,i})^2}{2\sigma_{m,i}^2}\right)
    \label{txl_transfer_function_g}
\end{equation}
This response is realised through an approximate double-sigmoid description whose product of two complementary sigmoid transitions (at $V_{TH}^{lower}$ and at $V_{TH}^{upper}$) produces a plateau-shaped window function, as shown in Fig. \ref{fig:figure_1}(d). For a specific $V_{ML}$, TXL array supply, and $R_{limiter}$, per TXL cell contribution resistance, the generated cell current is:
\begin{equation}
    I_{m,i}^{cell} = \frac{V_{ML}}{R_{limiter}} \cdot g_{m,i}(x_i)
    \label{i_cell_through_r_limiter}
\end{equation}
and the matchline current is accumulated via Kirchhoff's Current Law (KCL) across all $D$ cells in the row, yielding the aggregate matchline signal:
\begin{equation}
    I^{ML}_m(\mathbf{x}) = \sum_{i=1}^{D} I_{m,i}^{cell}(x_i)
\end{equation}
This summation is performed naturally through Kirchhoff's current law due to the interconnection of all TXL cells to a common node for each prototype. Due to matchline aggregation operation via accumulation of cell currents through KCL, the array computes an additive, feature-wise similarity score rather than the product of per-feature Gaussian terms that a conventional multi-dimensional Gaussian RBF would use. For the prototype ranking we are using the score as calculated by the total current accumulation per row. \par
For each TXL cell separately, matchline current is monotonically related to the feature-wise Gaussian response. At the TXL array level, the accumulated matchline current is an additive similarity score. An argmax function can be applied over matchlines to identifies the best match $m^* = \arg\max( I^{ML}_m(\mathbf{x}))$ \cite{rbf_9t4r_iscas}. Based on the current output of the TXL array the normalised similarity $\hat{\mathcal{N}}_m = I_m^{ML}/I_{max}^{ML} \in (0, 1]$ can be calculated, which provides a normalised hardware-related similarity score and reliability metric on the classification results. The normalised similarity is used to identify whether a input query is reliably in-distribution, unreliably in-distribution as outlier sample, or unknown out-of-distribution. The normalization reference should be interpreted as the calibrated full-match matchline response of the hardware, not the maximum matchline response among prototypes for the current query. \par
\subsection*{Prototypical Knowledge Adaptation} \label{sec:adaptation_methods}
The two boundary thresholds per prototype, $\tau_{IDO}$ (in-distribution outlier threshold) and $\tau_{OOD}$ (out-of-distribution threshold), are derived from the inverse cumulative distribution function of the dataset distribution at selected confidence levels $P_{IDO}$ and $P_{OOD}$, respectively. The thresholds are calculated heuristically based on the statistical analysis of the training dataset during the initial offline training and can be mapped in hardware as the reference signals of the sense amplifiers. For the cases of online adaptation, the streaming test data are used as either part of the training dataset, to calibrate existing prototypes, or as standalone training dataset for the case of adding a new prototype. The update rule for the mean and variance of a TXL cell follows the Eq. \ref{mean_update} and Eq. \ref{variance_update}, respectively. The plasticity factor $\eta \in [0,1]$ controls how the rate that the prototype tracks new observations when adaptation is applied. \par
\begin{equation}
    \boldsymbol{\mu}_{m^*}^{(t+1)} = (1-\eta)\,\boldsymbol{\mu}_{m^*}^{(t)} + \eta\,\mathbf{x}
    \label{mean_update}
\end{equation}
\begin{equation}
    \sigma_{m^*,i}^{2\,(t+1)} = (1-\eta)\,\sigma_{m^*,i}^{2\,(t)} + \eta\,(x_i - \mu_{m^*,i}^{(t+1)})^2
    \label{variance_update}
\end{equation}
Through equations \ref{mean_update} and \ref{variance_update}, the placement and width of each TXL cell's matching window in the row of interest is updated. When a sample is classified as OOD, TXL peripheral circuitry routes it to an OOD buffer $\mathcal{B}_{OOD}$. The OOD buffer mitigates issues arising from transient noise since a single anomalous input should be insufficient to justify allocating a new class prototype in many applications. Once the buffer reaches an arbitrary minimum size, the system evaluates whether the accumulated samples form a coherent cluster by measuring their internal prototype variance. A few-shot learning scheme is applied to generate and store the new prototype. The new prototype mean and variance are computed directly from the buffer as sample averages, and a new matchline is programmed with the corresponding RRAM resistance values. The array simply grows by one row and no existing prototype is overwritten or modified. Limitations on that total variance of the prototype can provide a simple method of identifying if the stored anomalous inputs constitute a new class. \par
\backmatter
\section*{Acknowledgements}\label{sec:acknowledgements}
The authors would like to acknowledge that this work was supported in part by the Engineering and Physical Sciences Research Council (EPSRC) Programme under Functional Oxide Reconfigurable Technologies (FORTE) Grant EP/R024642/2, the EPSRC ProSensing Project (Grant No. EP/Y030176/1) and the RAEng Chair in Emerging Technologies under Grant CiET1819/2/93. 
\section*{Data Availability}\label{sec:data_availability}
The relevant data for this work can be found in (TBC).
\section*{Code Availability}\label{sec:code_availability}
The relevant code for this work can be found by contacting the authors of this work. The authors can provide the relevant code used for the main simulations upon reasonable requests.
\section*{Author Contribution Statement}\label{sec:auth_contr_statement}
GP did the simulations in SPICE-Spectre and Python, designed and tested the circuits in Cadence Virtuoso, developed and tested the high-level behavioural model in Python, and contributed to the manuscript. SA and TP worked on the concept and contributed to the manuscript. 
\section*{Competing Interest Statement}\label{sec:comp_inter_statement}
The authors would like to state that they have no competing interests.
\bibliography{References_v2026_updated_e6aGP}
\section*{Supplementary Material \& Extended Data} \label{sec:supp_material}
In supplementary section \ref{TXL_Python_Supp}, the high-level Python code modelling the behaviour of TXL array is shared. The Python TXL model is used to test the expected behaviour of the TXL behaviour. This high-level behavioural model is used to provide additional Python simulation results for the artificial symbol dataset used for the SPICE-Spectre results shown in the results section. Additionally, the same model is used to provide an alternative approach to showcasing the inference, outlier and out-of-distribution adaptation of the TXL array through the well-documented MNIST dataset. \par 
In supplementary section \ref{TXL_Symbols_Supp}, additional simulation results for the example artificial symbol dataset. These additional simulation are performed in Python using a high-level behavioural TXL model that test the clustering capabilities of the TXL array and provides further examples on how the TXL operates in performing inference via metric-based distance calculations and backpropagation-free adaptation of the prototypical memory knowledge. \par
In supplementary sections \ref{TXL_MNIST_Supp}, additional simulation results for benchmarking the Python-based high-level behavioural TXL model using MNIST dataset. The example results are based on a similar operation verification method as with the artificial symbol dataset. \par
In supplementary section \ref{TXL_Matchline_Supp}, additional information are showcased on linear analogue charging effect of the TXL matchline by increasingly activating TXL cell contributions. The simulations were performed in Cadence Virtuoso environment using a commercially available $\SI{180}{\nano\meter}$ and $\SI{5}{\volt}$ devices to satisfy our in-house RRAM device technology electroforming and analogue/multi-bit programming requirements \cite{9t4r_tcasi, rbf_9t4r_iscas}. \par
\subsection{Approximate TXL Python Model} \label{TXL_Python_Supp}
A high-level approximation of the behavioural response of the TXL array is used to quickly prototype Python-based examples of simple proof-of-concept applications of TXL in classification operations. The main operation modelled is the distance calculation of an input against the store templates in the TXL array. The TXL inference calls the distance calculation function and performs an integrated best match policy through a winner take all emulation. 
The simplified high-level TXL model used for the proof-of-concept Python simulation is based on the pseudo-code in the following algorithm that clearly illustrates the high-level concepts followed.

\begin{algorithm}
\caption{TXL-ACAM --- Hardware Constant Derivation}
\label{alg:setup}
\begin{algorithmic}[1]
\Require Hardware dictionary $\mathcal{H} = \{V_{DD}, V_{tn}, V_{tp}, \beta_n, \beta_p, I_s, R_B, V_{ML}, R_{lim}, R_{min}, R_{max}, P_{IDO}, P_{OOD}\}$
\Ensure Derived constants $(k_r, V_{TH0}, A)$
\State $k_r \gets \sqrt{\beta_p / \beta_n}$
\State $V_{TH0} \gets \dfrac{V_{tn} + k_r(V_{DD} - V_{tp})}{1 + k_r}$
\State $A \gets \dfrac{1 + k_r}{I_s \, k_r}$
\State \Return $(k_r, V_{TH0}, A)$
\end{algorithmic}
\end{algorithm}

\begin{algorithm}
\caption{TXL-ACAM --- Cell Programming (Encode)}
\label{alg:encode}
\begin{algorithmic}[1]
\Require Target statistics $(\mu_i, \sigma_i)$ for feature $i$; constants $(k_r, V_{TH0}, A)$; bounds $[R_{min}, R_{max}]$
\Ensure Programmed resistance pair $(R_{M1,i}, R_{M2,i})$
\Function{Encode}{$\mu_i, \sigma_i$}
    \State $R_{\text{base}} \gets R_B / k_r$
    \State $R_{M1} \gets R_{\text{base}} - A\,(\mu_i - \sigma_i - V_{TH0})$ \Comment{lower threshold branch}
    \State $R_{M2} \gets R_{\text{base}} - A\,(\mu_i + \sigma_i - V_{TH0})$ \Comment{upper threshold branch}
    \State $R_{M1} \gets \operatorname{clip}(R_{M1}, R_{min}, R_{max})$
    \State $R_{M2} \gets \operatorname{clip}(R_{M2}, R_{min}, R_{max})$
    \State \Return $(R_{M1}, R_{M2})$
\EndFunction
\end{algorithmic}
\end{algorithm}

\begin{algorithm}
\caption{TXL-ACAM --- Inference (Row-wise Similarity Search and Decision)}
\label{alg:inference}
\begin{algorithmic}[1]
\Require Query vector $\mathbf{x} \in \mathbb{R}^D$; $M$ prototype rows, each with programmed $\{(R_{M1,m,i}, R_{M2,m,i})\}_{i=1}^{D}$; constants $(k_r, V_{TH0})$; $\mathcal{H}$

\Statex \textbf{Stage 1: per-row decode}
\For{each row $m = 1, \dots, M$}
    \For{each cell $i = 1, \dots, D$}
        \State $V_{lo} \gets V_{TH0} + \dfrac{I_s\,(R_B - k_r R_{M1,m,i})}{1+k_r}$
        \State $V_{hi} \gets V_{TH0} + \dfrac{I_s\,(R_B - k_r R_{M2,m,i})}{1+k_r}$
        \State $\hat\mu_{m,i} \gets (V_{hi} + V_{lo})/2 \, ; \quad \hat\sigma_{m,i} \gets \max\!\big((V_{hi}-V_{lo})/2,\ \epsilon\big)$
    \EndFor
\EndFor

\Statex \textbf{Stage 2: per-row similarity (KCL matchline sum, Eq.~10)}
\For{each row $m = 1, \dots, M$}
    \State $I_{ML,m} \gets \dfrac{V_{ML}}{R_{lim}} \displaystyle\sum_{i=1}^{D} \exp\!\left(-\dfrac{(x_i - \hat\mu_{m,i})^2}{2\hat\sigma_{m,i}^2}\right)$
    \State $\hat N_m \gets \operatorname{clip}\!\left(\dfrac{I_{ML,m}}{D \cdot V_{ML}/R_{lim}},\ 0,\ 1\right)$
\EndFor

\Statex \textbf{Stage 3: winner selection}
\State $m^* \gets \arg\max_m \hat N_m$

\Statex \textbf{Stage 4: reliability gating}
\State $d^2_{m^*} \gets \displaystyle\sum_{i=1}^{D} \left(\dfrac{x_i - \hat\mu_{m^*,i}}{\hat\sigma_{m^*,i}}\right)^{2}$
\State $\tau_{IDO} \gets F^{-1}_{\chi^2_D}(P_{IDO})\, ; \quad \tau_{OOD} \gets F^{-1}_{\chi^2_D}(P_{OOD})$
\If{$d^2_{m^*} \le \tau_{IDO}$}
    \State \textbf{zone} $\gets$ \textsc{Reliable}
\ElsIf{$d^2_{m^*} \le \tau_{OOD}$}
    \State \textbf{zone} $\gets$ \textsc{Unreliable (IDO)}
\Else
    \State \textbf{zone} $\gets$ \textsc{OOD}
\EndIf
\State \Return $(m^*, \hat N_{m^*}, d^2_{m^*}, \textbf{zone})$
\end{algorithmic}
\end{algorithm}

\begin{algorithm}
\caption{TXL-ACAM --- Prototype Adaptation}
\label{alg:adapt}
\begin{algorithmic}[1]
\Require Winning row $m^*$ with current $(\mu^{(t)}_{i}, \sigma^{(t)}_{i})_{i=1}^{D}$; new sample $\mathbf{x}$; plasticity $\eta \in [0,1]$
\Ensure Re-programmed $(R_{M1,i}, R_{M2,i})_{i=1}^{D}$ for row $m^*$
\If{\textbf{zone} $=$ \textsc{Unreliable (IDO)}} \Comment{existing-class update}
    \For{each cell $i = 1, \dots, D$}
        \State $\mu^{(t+1)}_i \gets (1-\eta)\,\mu^{(t)}_i + \eta\, x_i$
        \State $\big(\sigma^{(t+1)}_i\big)^2 \gets (1-\eta)\,\big(\sigma^{(t)}_i\big)^2 + \eta\,\big(x_i - \mu^{(t+1)}_i\big)^2$
        \State $(R_{M1,i}, R_{M2,i}) \gets \Call{Encode}{\mu^{(t+1)}_i, \sigma^{(t+1)}_i}$ \Comment{Algorithm~\ref{alg:encode}}
    \EndFor
\ElsIf{\textbf{zone} $=$ \textsc{OOD} \textbf{and} allocation triggered}\Comment{new-class allocation}
    \State Accumulate $n$ OOD-buffered samples $\{\mathbf{x}^{(1)}, \dots, \mathbf{x}^{(n)}\}$
    \For{each cell $i = 1, \dots, D$}
        \State $\mu_i \gets \frac{1}{n}\sum_{j=1}^n x^{(j)}_i \, ; \quad \sigma_i \gets \operatorname{std}_j\!\big(x^{(j)}_i\big)$
        \State $(R_{M1,i}, R_{M2,i}) \gets \Call{Encode}{\mu_i, \sigma_i}$
    \EndFor
    \State Append as a new row $m^* \gets M{+}1$; existing rows unchanged
\EndIf
\end{algorithmic}
\end{algorithm}

\subsection{TXL Example Application on Artificial Symbols Dataset} \label{TXL_Symbols_Supp}
\begin{figure*}[h!]
	\centering
	\includegraphics[width=\linewidth]{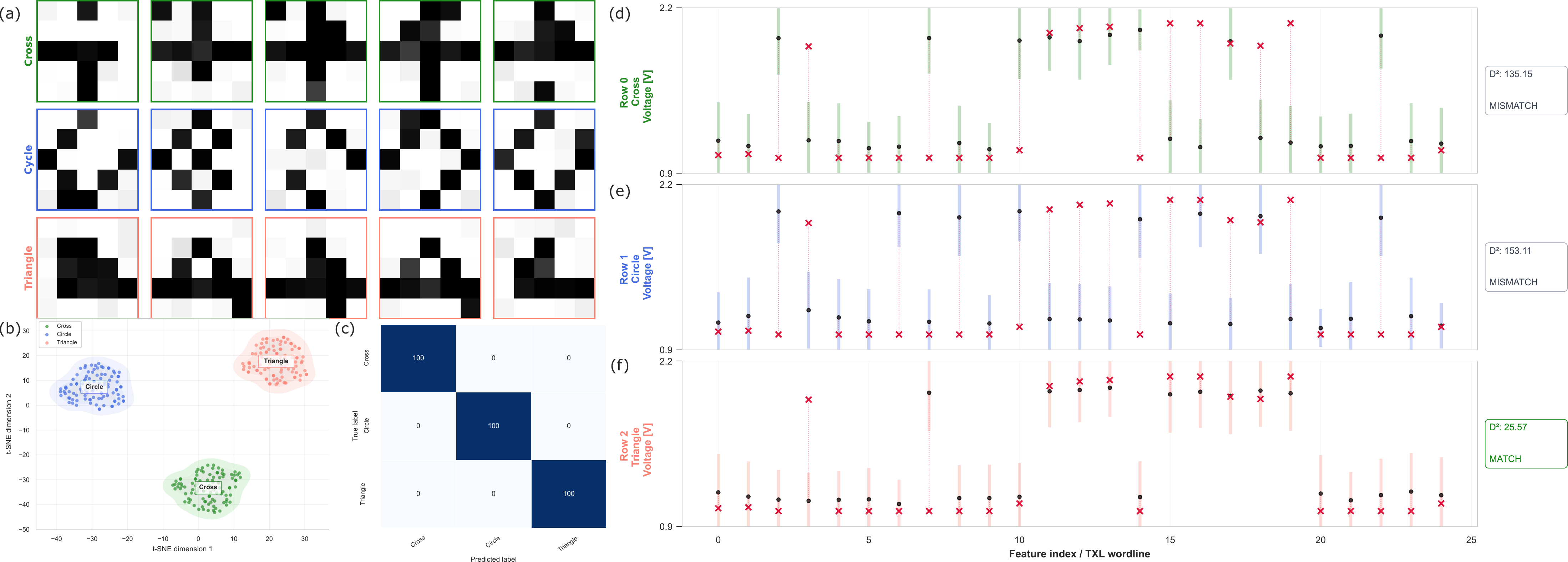}
	\caption{\textbf{TXL Python Test on Test Artificial Symbols Dataset}: \textbf{(a)}  Example samples of the three main classes (crosses, circles, and triangles). Each sample if an artificially generated $5\times5$ image. \textbf{(b)} Cluster plot after the clustering has been applied to generate the representative prototypes in TXL array. \textbf{(c)} Confusion table for a test inference for the TXL with the extracted templates. \textbf{(d)} Example query of a triangle sample through the TXL array. The TXL array searches and compares the query input with all stored prototypes in a single step. The best match is calculated and then compared with the IDO and OOD threshold.}
	\label{fig:figure_1_symbols}
\end{figure*}

\begin{figure*}[h!]
	\centering
	\includegraphics[width=\linewidth]{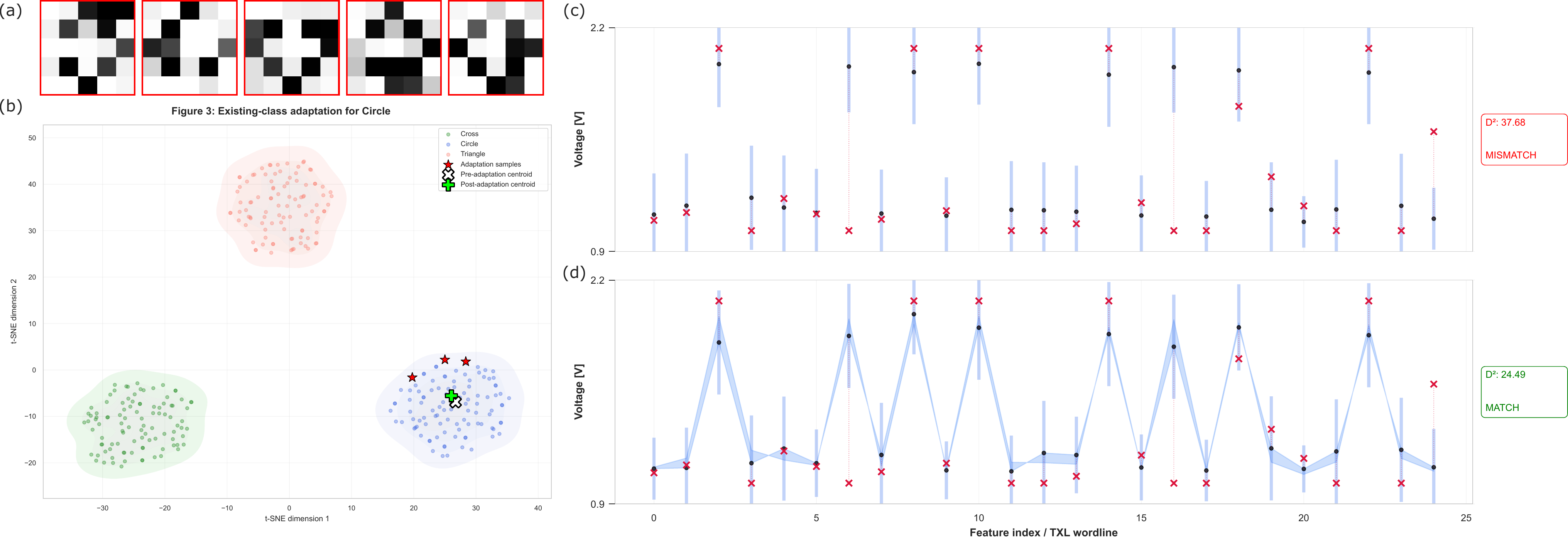}
	\caption{\textbf{TXL Adaptation for Domain Shift in Artificial Symbols}: \textbf{(a)} Example set of noisy outlier examples of the circle class that are used to adapt the existing circle prototype to recognise the outliers. \textbf{(b)} The updated cluster plot with the centroid of the circle class being adapted towards the direction of the outlier samples to better represent those as well. \textbf{(c)} The pre-adaptation prototype of the circle class that fails to correctly identify one of the outlier samples. \textbf{(d)} The post-adaptation prototype of the circle class that correctly identifies the same outlier sample. The changes to the densitometric representation of circle class are shown one top of the prototype representation. }
	\label{fig:figure_2_symbols}
\end{figure*}

\begin{figure*}[h!]
	\centering
	\includegraphics[width=\linewidth]{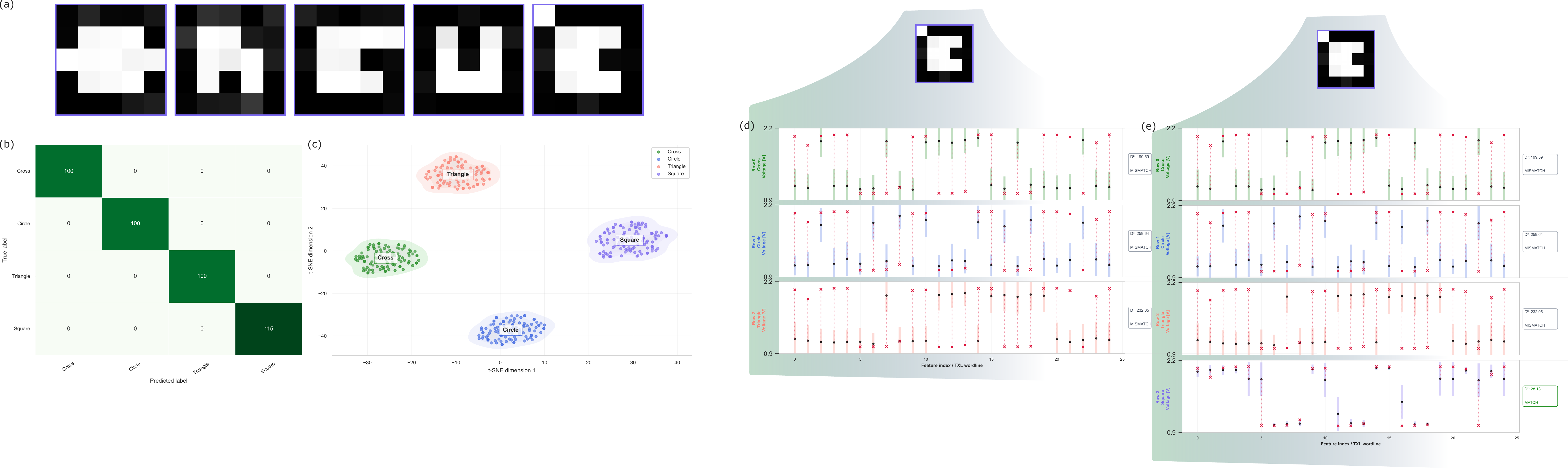}
	\caption{\textbf{TXL Adaptation for a New Symbol}: \textbf{(a)} Example samples from a new out-of-distribution class (rectangles) that is used in a supervised online learning scheme to showcase easy on-the-fly adaptation of the TXL array with new prototypes without requiring backpropagation. \textbf{(b)} Updated confusion table after the few-shot learning showcase expansion of the TXL with an additional class without showing any catastrophic forgetting. The prototype generation and updated inference is performed through a fully explainable and interpretable method by applying clustering to the novel class samples. \textbf{(c)} Updated cluster plot that shows the old classes alongside the new one. The cluster plot is represented through t-SNE encoding. \textbf{(d)} The pre-adaptation TXL memory inference where a sample from the new rectangle class is queried to the existing prototypes with all row mismatching and showing big distance from encoded knowledge. \textbf{(e)} The post-adaptation TXL memory inference where the same sample is queried the the updated list of prototypes with all prior prototypes responded with the same mismatch while the newly generated rectangle prototype providing a match response for the newly added prototype.}
	\label{fig:figure_3_symbols}
\end{figure*}

Supplementary information for the software deployment of TXL behavioural model in classification tasks for a generated artificial dataset using noisy symbols. In Fig. \ref{fig:figure_1_symbols}, showcase the initial clustering of 2 classes of the artificial symbol dataset. The figure showcase input samples from the different classes used, clustering plot through t-SNE representation, a confusion table and the generated and stored prototype per class. In Fig. \ref{fig:figure_2_symbols}, an adaptation of existing prototype from one of the initially known class is showcased. The figure shows the noisy samples used for the few-shot learning test, the cluster plot that showcase the movement of the class centroid to adapt to the new outlier class samples and the pre-adaptation and post-adaptation class prototype. As shown, the pre-adaptation prototype fails to match with an outlier sample while for the same sample the post-adaptation prototype is matching. In Fig. \ref{fig:figure_3_symbols}, an adaptation of the TXL array when encountering a new type of symbols, thus new class, through few-shot learning is shown. The figure showcase example samples from the new class that are used to generate the new prototype in TXL array, the updated confusion table from a test inference when the new class is included, the updated cluster plot, as well as the state of the TXL array before and after adaptation when queried with a new class sample. As seen prior to adaptation with the new prototype the novel class cannot be classified due to large distance from all existing prototypes. After the adaptation with the new prototype generated by few-shot learning method, the novel input can be classified. \par
\subsection{TXL Example Application on MNIST Dataset} \label{TXL_MNIST_Supp}
\begin{figure*}[h!]
	\centering
	\includegraphics[width=\linewidth]{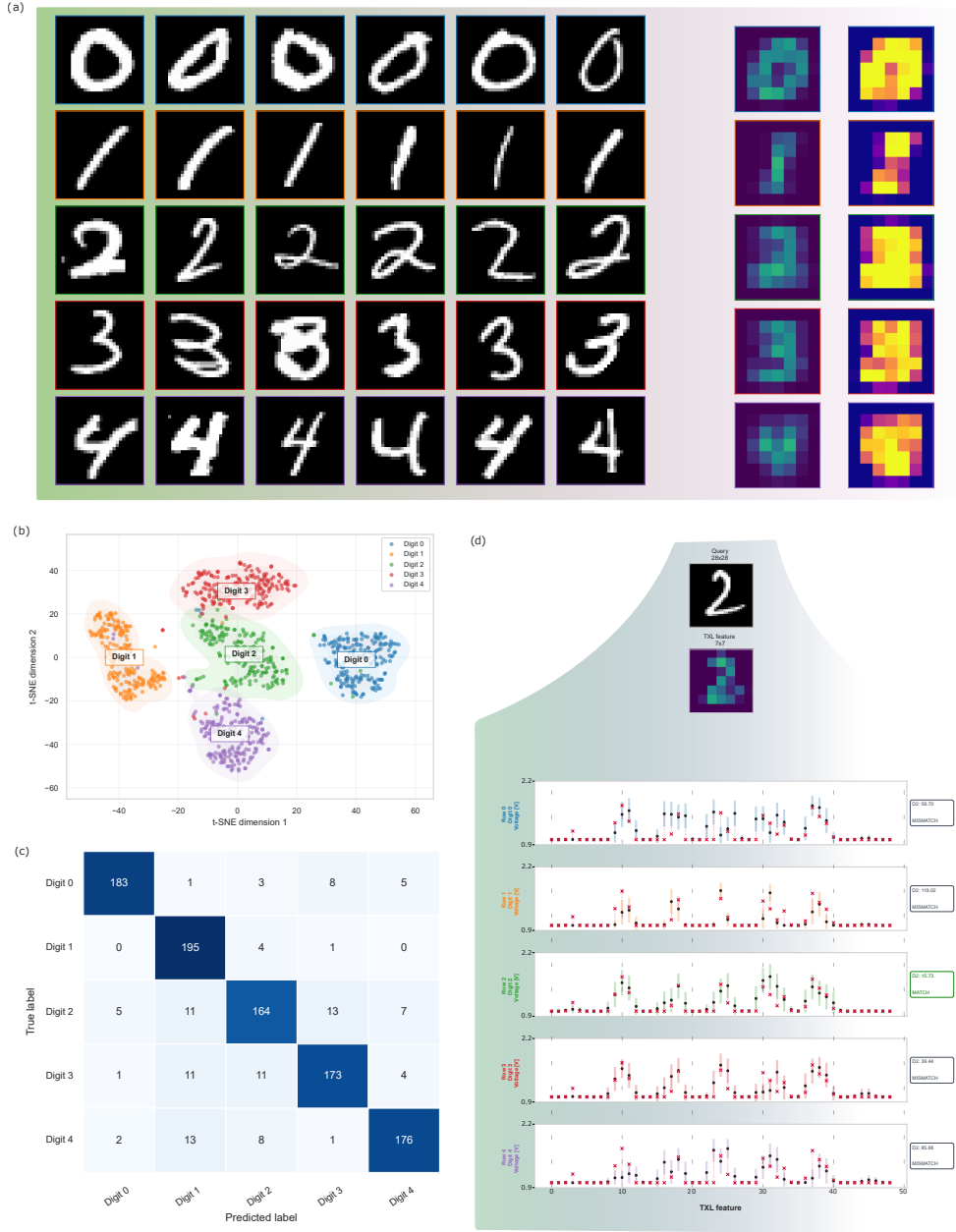}
	\caption{\textbf{TXL Python Test on MNIST Dataset}: \textbf{(a)} Downsampled $7\times7$ MNIST input for easier compatibility with the query length of the TXL9v1 prototype IC array. The images are downsampled to $7\times7$ images, thus 49-dimensional input vectors. From the downsampled MNIST images, the centroid and variance prototypical maps are generated and then programmed to the appropriate entry matchline of the TXL array. \textbf{(b)} Clustering plot for the initial classes the network is performing a supervised clustering to form the initial set of representative prototypes. \textbf{(c)} Confusion matrix for inference test on the first 5 classes of MNIST. \textbf{(d)} Example similarity search operation for digit 2.}
	\label{fig:figure_1_mnist}
\end{figure*}

\begin{figure*}[h!]
	\centering
	\includegraphics[width=\linewidth]{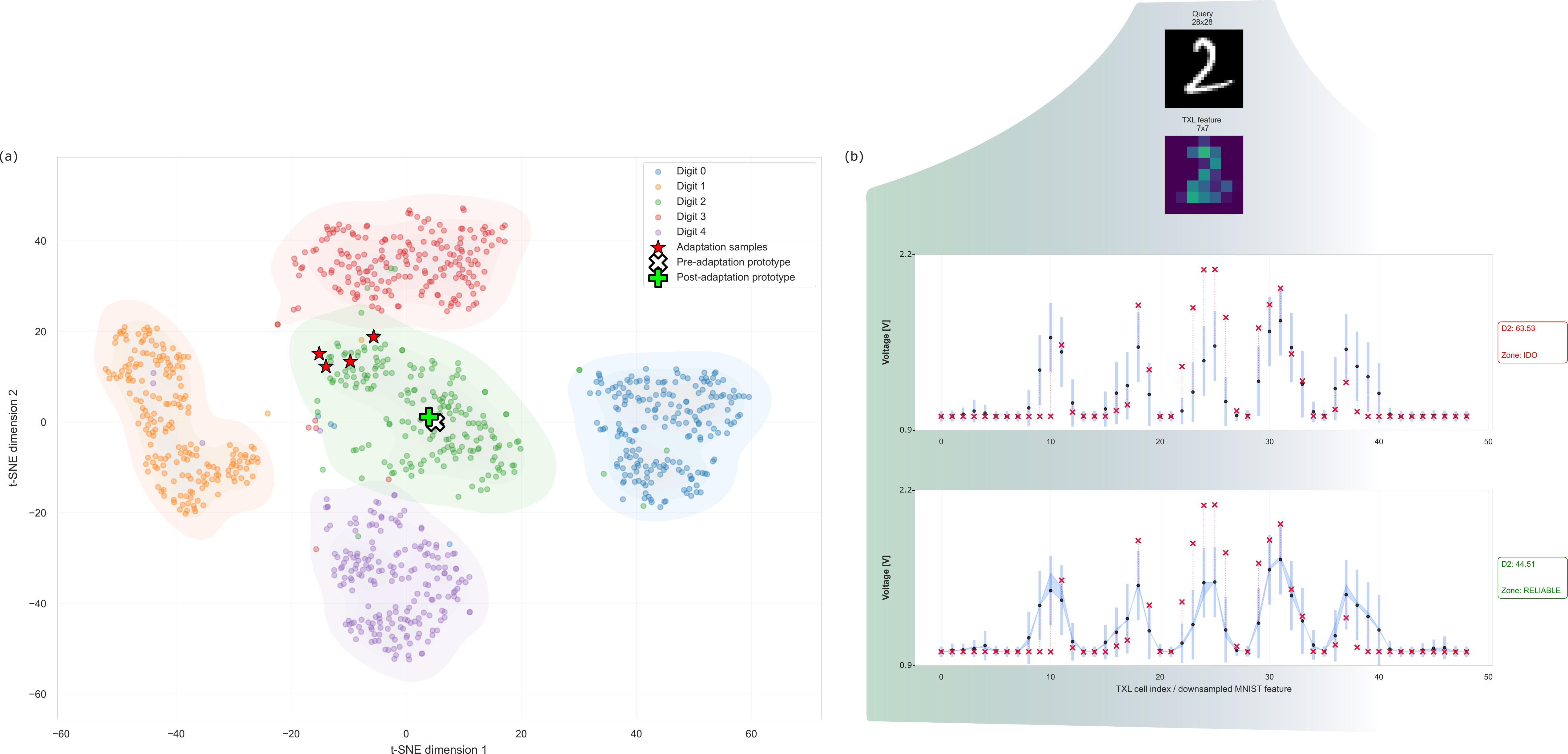}
	\caption{\textbf{TXL Adaptation for Domain Shift in MNIST}: \textbf{(a)} Clustering plot with outliers for the class 0 and small prototype adaptation to capture the added cases. \textbf{(b)} On top the example outlier input is shown in full resolution and then downsampled. In the upper trace, the pre-adaptation similarity search results in mismatch for the relevant prototype. In the lower trace, the post-adaptation similarity search results in match after a limited adaptation of the relevant prototype.}
	\label{fig:figure_2_mnist}
\end{figure*}

\begin{figure*}[h!]
	\centering
	\includegraphics[width=\linewidth]{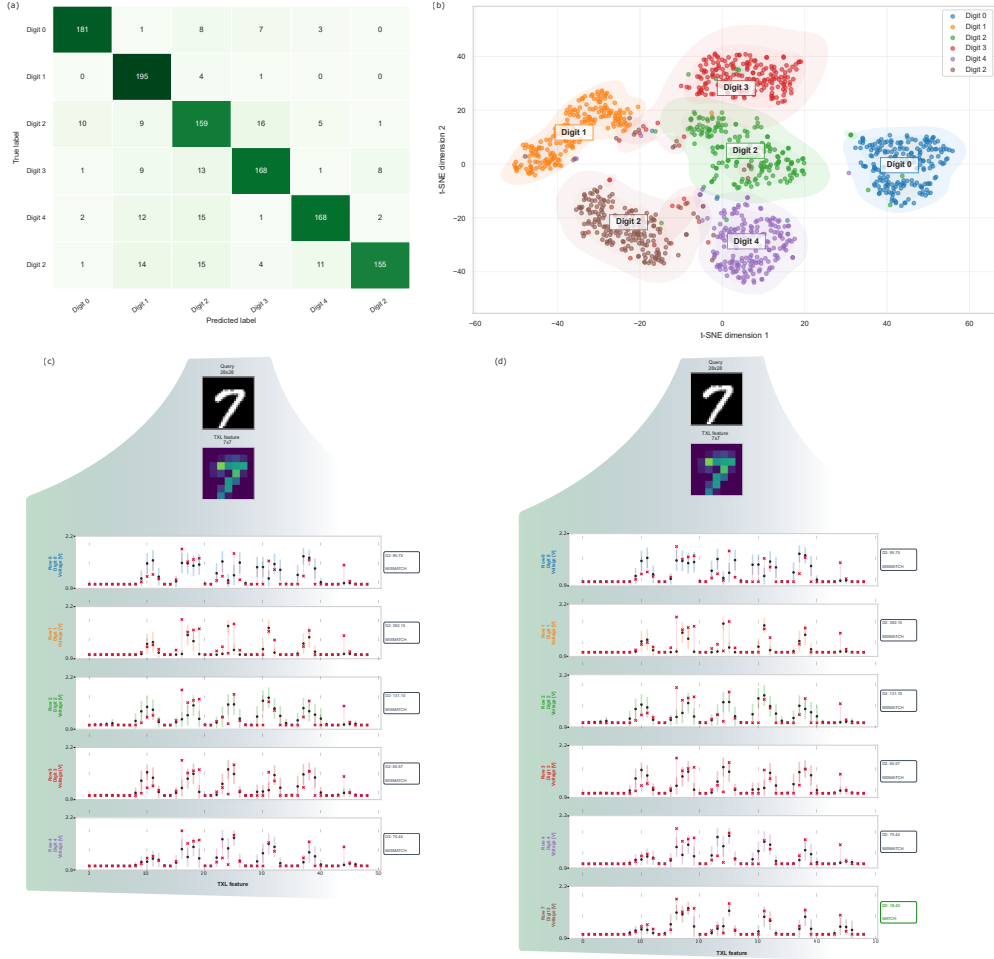}
	\caption{\textbf{TXL Adaptation for New MNIST Class}: \textbf{(a)} Updated confusion table with the additional class 7 prototype included and tested. \textbf{(b)} Updated clustering plot with 6 class representations. \textbf{(c)} Status of TXL prototype knowledge before the online learning of class 7. \textbf{(d)} Status of TXL prototype knowledge after learning and encoding class 7 prototype.}
	\label{fig:figure_3_mnist}
\end{figure*}

Supplementary information for the software deployment of TXL behavioural model in classification tasks for MNIST dataset. In Fig. \ref{fig:figure_1_mnist}, the initial supervised clustering is providing the first set of known MNIST prototypes is exhibited. In Fig. \ref{fig:figure_2_mnist}, a few-shot learning example of adapting a known class to recognise outlier samples is shown. In Fig. \ref{fig:figure_3_mnist}, a few-shot learning example of generating a new prototype for a new class is shown.
\subsection{TXL Matchline SPICE Simulation in Cadence Virtuoso} \label{TXL_Matchline_Supp}
\begin{figure*}[h!]
	\centering
	\includegraphics[width=\linewidth]{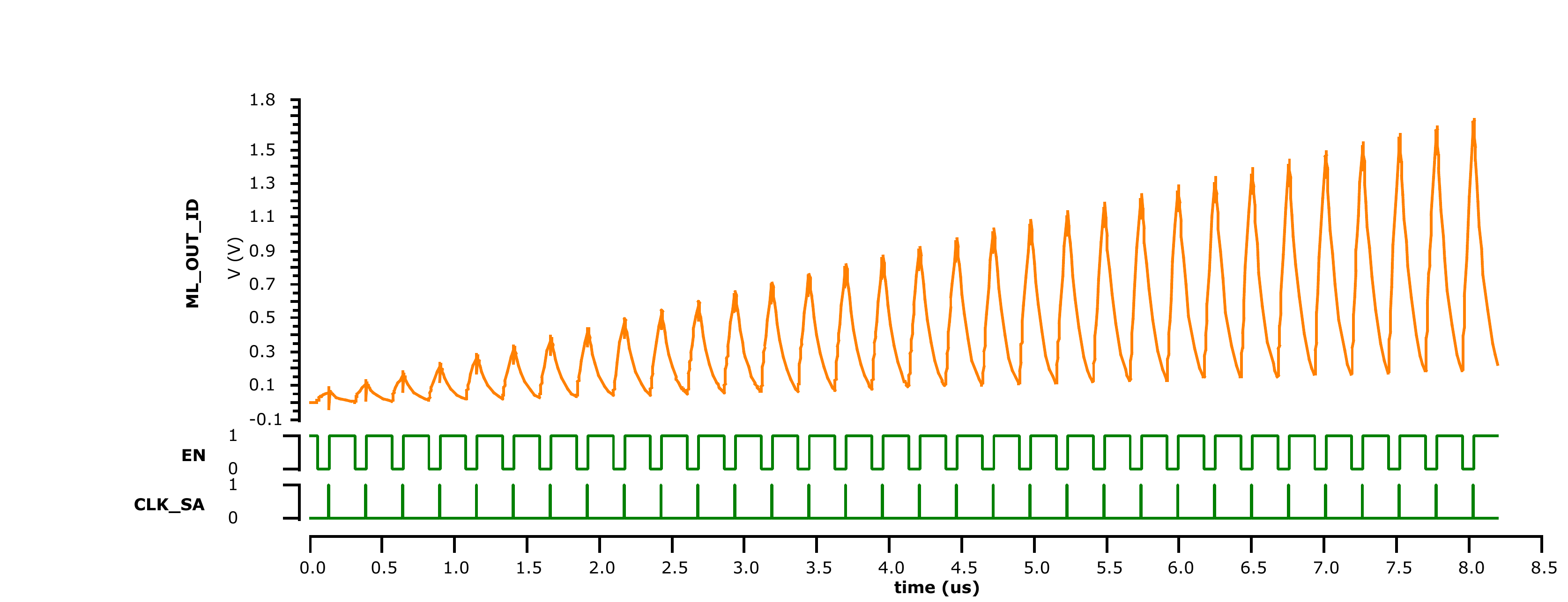}
	\caption{\textbf{Linear Matchline (ML) Charging Behavioural Characterisation}: In the upper (orange) the analogue matchline output is captured through SPICE-Spectre simulations. The signal showcase charging and discharging patterns for a matchline that for each step another TXL cell is generating a match charging event, thus the feature input for this specific TXL cell falls within its matching range. The lower (green) digital controls is for the enable signal (EN) and the sense amp clock (CLK\_SA) which both are used to determine the charge-discharge cycle in the TXL matchline showcased operation. The simulation is a transient analysis using commercially available $\SI{180}{\nano\meter}$ $\SI{5}{\volt}$ MOSFET components and in-house model for our RRAM devices in Cadence Virtuoso environment. }
	\label{fig:figure_1_matchline}
\end{figure*}

Supplementary information for SPICE-Spectre simulation performed in Cadence Virtuoso environment to test the circuits and systems level behavioural response of TXL cells and arrays. In Fig. \ref{fig:figure_1_matchline}, an example SPICE-Spectre simulation of a row in TXL array that uses a vector stimulus input that gradually activates the TXL cells. The simulation showcase good linearity of charge accumulation on the matchline that is based on the Kirchhoff's second law natural computing.
\end{document}